\documentclass[12pt]{iopart}
\usepackage{amssymb}
\usepackage{graphicx}

\usepackage{bm}
\newcommand{\be}{\begin{equation}}
\newcommand{\ee}{\end{equation}}

\newcommand{\beq}{\begin{eqnarray}}
\newcommand{\eeq}{\end{eqnarray}}

\def\eq#1{(\ref{#1})}
\def\H1{\widehat{H}_1}

\begin{document}

\title[]{Topologically protected strongly correlated states of photons}

\author{Matous Ringel$^{1,2}$, Mikhail Pletyukhov$^3$,  Vladimir Gritsev$^{4}$ }

\address{
$^1$ Department of Physics, University of
Fribourg, Chemin du Mus\'ee 3, 1700 Fribourg, Switzerland\\
$^2$ Swiss Life Asset Management AG, General-Guisan-Quai 40, P.O.Box 2831,
8022 Z\"{u}rich, Switzerland\\
$^3$Institute for Theory of Statistical Physics and JARA -- Fundamentals of Future Information Technology, 
RWTH Aachen, 52056 Aachen, Germany \\
$^4$ Institute for Theoretical Physics, University of Amsterdam, Science Park 904,
Postbus 94485, 1098 XH Amsterdam, The Netherlands
}
\ead{pletmikh@physik.rwth-aachen.de; matous.ringel@gmail.com; V.Gritsev@uva.nl}
\begin{abstract}
Hybrid photonic nanostructures allow the engineering of novel interesting states of light.  One recent
example is topological photonic crystals where a nontrivial Berry phase of the photonic band
structure gives rise to topologically protected unidirectionally-propagating (chiral) edge states of
photons.  Here we demonstrate that by coupling an array of emitters to the chiral photonic edge
state one can create strongly correlated states of photons in a highly controllable
way. These are topologically protected and have a number of remarkable universal
properties: The outcome of scattering does not depend on the positions of emitters and is
given only by universal numbers, the zeroes of Laguerre polynomials; two-photon correlation 
functions manifest a well-pronounced even-odd effect with respect to the number of emitters, and 
the result of scattering is robust with respect to fluctuations in the emitters' transition frequencies. 
\end{abstract}

\maketitle

\section{Introduction}

Light-matter interaction surrounds us everywhere in nature, and it has played a tremendous role in the development
of current technology. Until recent decades it was sufficient to deal with this interaction on
average, with many photons and many atoms involved. However, the increasing miniaturization of basic
constituents towards nanoscale is a common trend in modern technology. Downscaling to single-atom
and/or single-photon levels promotes some traditionally classical research areas into the quantum
realm \cite{vahala04,fleischhauer05,polzik,carusotto}. A quantum control over 
the light-matter interaction will eventually become a vital ingredient of emerging quantum devices, and 
it is equally important for the development of several related fields, including communication, signal
processing, ultrafast optics, optomechanical cooling, imaging and spectroscopy, and
quantum information \cite{PhotQTech}. However, the
efficient manipulation and control requires a relatively strong
interaction at the level of a single atom or a single photon
\cite{res-fluor,hwang,abdumalikov,1-photon-rev,volz,1-photon-GHz,Yao}. This presents a significant
challenge, since the typical interaction scale for individual particles is given
by the smallness of the QED coupling constant $\alpha\approx 1/137$. Two possible ways to overcome this natural
limitation, and to increase the effects of correlations, are either to use artificial materials and
devices, or to resort to many-body effects to produce nonlinearities.

\begin{figure}[h]
\includegraphics[width=\textwidth]{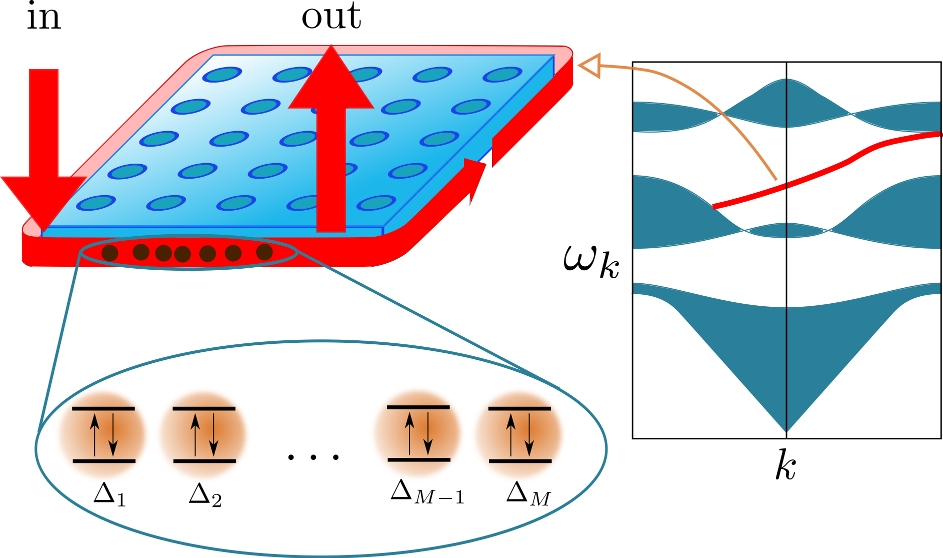}
\caption{{\bf The chiral 
    edge state of a topological photonic crystal.} A schematic view of the proposed setup to generate
    strongly correlated states of photons: The topological photonic insulator (left) 
    possesses a topologically nontrivial band structure (right). If the total Chern
    number of the bands below the gap is 1, a boundary state inside the gap is formed (its dispersion is denoted by the red line in the right
    picture). This state corresponds to the chiral edge state of unidirectionally propagating
    photons (thick red line at the boundary of the photonic crystal). We suggest to embed emitters
   of an arbitrary level structure (here with two levels) into the edge channel.
    A population of the edge states  by an external few-photon source (in-state) and their propagation through the array of emitters create strongly correlated
    outgoing photonic states (out-state).}
 \label{system}
\end{figure}

Recent experimental progress in fabricating few-photon sources coupled to one-dimensional (1D) transmission
lines \cite{schuster,dayan,wallraff,aoki,astafiev,1-photon,reinhard,hoi} opens an avenue for
creating and manipulating strongly correlated states of photons. It has also triggered a significant
number of theoretical studies \cite{1dscat-shenfan,1dscat-shi,shen-fan-LS,1dscat-chang,WS,transM2,Roy}.  
Our experience from condensed matter and atomic physics teaches us that combined effects of 
reduced dimensionality and interactions can often effectively enhance 
correlations, and eventually lead to new collective states of matter with properties which are
very different from those of the individual particles (e.g., Luttinger liquids in 1D).  
This insight is one of the driving forces behind the quest for novel correlated states of photons
\cite{corr-states-hartmann,corr-states-greentree,corr-states-chang}.

A different class of collective phenomena is the topological
insulating/superconducting state of matter. 
This state has been recently experimentally realized
for electrons \cite{top-iso-exp} and extensively studied theoretically~\cite{rev-top-ins}.
The main signature of  topological properties in the
band structure of bulk  materials
is an existence of {\it edge states}, which are insensitive to local perturbations, impurities, or geometrical imperfections.
In some -- chiral -- instances edge states propagate unidirectionally.
Recently, the existence of chiral states in photonic systems has been theoretically predicted \cite{Haldane_Raghu} and  experimentally observed \cite{MIT,top-light-exp,FYF,optical,haf1,floquet}. 
These edge states represent an optical
analogue of the quantum Hall edge states, and they are
 characterized by a nontrivial Chern number \cite{top-ins-class}.

In this paper, we suggest the use of one-dimensional unidirectional edge states  
as a robust platform for the controllable generation of strongly correlated states of photons.
To achieve this goal, we couple an array of emitters to a chiral edge state, see Fig.~(\ref{system}). Experimentally available novel hybrid photonic systems \cite{Cox} can be used to realize this setup. Photons in the
edge channel, populated by an external few-photon source, interact with an ensemble of emitters and
produce outgoing states with robust, controllable, and universal properties originating from the topological nature of the edge state. 
In particular, we find that the outgoing photonic wavefunction does not contain
any information about the positions of emitters; its nodes are rather determined by universal numbers -- the zeroes of Laguerre polynomials. An initial single-photon wavepacket is fragmented into pieces between these nodes.
In the case of two-photon scattering, we observe a clearly pronounced parity effect with respect to the number of emitters, which manifests in a transition from photons' bunching to antibunching as one changes the parity of the number of emitters from even to odd. We also show that the observed properties are robust with respect to fluctuations in the emitters' transition frequencies. The proposed setup can be experimentally realized in the GHz and optical domains with existing nanophotonic elements.

\section{1D edge modes interacting with emitters}

We consider topologically protected states propagating unidirectionally at the edge of the topological photonic crystal. We propose to embed an array of two-level (in general -- multi-level) 
systems -- called emitters -- into the chiral edge channel. We
assume that  the transition frequencies of emitters are commensurate with the frequency of the propagating chiral
mode of light. Assuming also that the excited emitter states mainly decay into the one-dimensional mode (with a decay rate $\Gamma_{\mathrm{1D}}$), we note that there are several sources of inevitable losses.  Namely, the excited emitter can decay into the continuum of three-dimensional modes in ambient space (with a decay rate $\Gamma_0$), into the bulk modes of a 2D photonic crystal (decay rate $\Gamma_{\mathrm{2D}}$), or to its impurity (bound) states (decay rate $\Gamma_{\mathrm{bs}}$). Here we note that the decay into the two-dimensional bulk modes of a photonic crystal is suppressed, $\Gamma_{\mathrm{2D}}=0$, since their density of states is zero in the bulk bandgap. We also assume that a 2D photonic crystal is clean enough to neglect the coupling of edge states to eventual mid-gap impurity states.

To generate well-defined chiral modes, it is necessary to minimize losses occurring at the rate $\Gamma_0$. The waveguide Purcell factor \cite{plasmons2}
\beq
\frac{\Gamma_{\mathrm{1D}}}{\Gamma_0} = \frac{3 }{\pi \epsilon_d^{3/2}} \frac{c}{v_g} \frac{A_0}{A_{\mathrm{eff}}}
\label{purcell}
\eeq
must then considerably exceed unity, which defines the strong coupling regime. In this expression $A_{0}=\lambda_{0}^{2}/4$ is the minimal cross-sectional area to confine the light in vacuum (for the light's wavelength $\lambda_{0}$); $\epsilon_{d}$ is the dielectric constant of the host medium where the emitter is embedded; $v_{g}$ is the group velocity of the propagating edge mode. The cross-sectional area $A_{\mathrm{eff}}$ of the effective confinement is estimated by $\sim b \xi$, where $b$ is the thickness of the slab, and $\xi \sim c \hbar/B$ is the localization depth of the edge mode, $B$ being the bulk bandgap. Thus, the waveguide Purcell factor \eq{purcell} can be enhanced in three different ways: 1) by reducing the thickness $b$ of the slab; 2) by increasing the bandgap $B$; and 3) by reducing the group velocity $v_g$.  All these methods to achieve the strong coupling regime have been successfully applied in plasmonic nanowires \cite{1dscat-chang,chang1,akimov} and in the Floquet photonic topological insulators made of helical waveguides \cite{floquet}.

Deep inside the bulk bandgap we can model the dispersion of the chiral edge mode by a linear dependence. In the following we measure all gauge-dependent energy scales from the spectrum linearization point. In the strong coupling regime we neglect losses into the three-dimensional continuum, and consider only the interaction of the edge mode with an array of $M$ emitters. Emitters are modeled by two-level systems with transition frequencies $\Delta_a$ and couplings $\sqrt{\kappa_a} \sim \sqrt{\Gamma_{1 \mathrm{D}}}$, and are placed at different positions $x_a$, which are separated from each by distances larger than the light's wavelength (an arrangement opposite to Dicke's \cite{dicke}). Thus, the Hamiltonian of our model reads
\beq
H&=&-i\int dx a^{\dag}(x)\partial_{x} a(x)
+\sum_{a=1}^{M}\Delta_{a}\left( S^{z}_{a}+\frac{1}{2} \right)\nonumber\\
&-& \sum_{a=1}^{M} \sqrt{\kappa_a} [S^{+}_{a} a(x_{a})+ a^{\dag}(x_{a})S^{-}_{a}],
\label{model}
\eeq
where we use units such that $v_g = \hbar = 1$. The chiral field operators $a^{\dag}(x), a(x)$ satisfy the standard commutation relations $[a(x),a^{\dag}(x')]=\delta(x-x')$. Transitions between emitters' states are described by the operators $S_a^{\pm}$, which satisfy the standard spin algebra $[S_a^{z}, S_{a'}^{\pm}]=\pm S_a^{\pm} \delta_{aa'}$, $[S_a^{+},S_{a'}^{-}]=2S_{a}^{z} \delta_{aa'}$. Spontaneous emission to other modes out of the one-dimensional waveguide is modeled by attributing an imaginary part $- i \kappa'/2 \equiv -i (\Gamma_0 + \Gamma_{\mathrm{bs}})/2$ to the transition frequencies $\Delta_a \to \Delta_a - i \kappa'/2$, in the spirit of the quantum jump picture \cite{carmichael}.

The problem is further specified by an initial state. Inspired by  experimental realizations of waveguides coupled to  emitters \cite{1-photon,reinhard}, we assume that the edge states are populated by an external
few-photon source, while emitters are initially prepared in the ground state. Injected photons
propagate unidirectionally and interact with an ensemble of
emitters, and after a sufficiently long time a stationary state is eventually established: Emitters typically relax back to the ground state, while the photonic wavefunction is modified. The evolution of the $N$-photon wavefunction is described in terms of the scattering matrix $S^{(N)}_{M;\{ k_i\},\{ k'_i\}} \equiv S^{(N)}_{M;\{ k_i\},\{ k'_i\}} (\{ \Delta_a \}, \{ \kappa_a \})$ depending of the sets of all transition frequencies $\Delta_a$ and couplings $\kappa_a$
\beq
\phi_{N, \mathrm{out}} (\{k_i\}) = \int d \{ k'_i\} S^{(N)}_{M;\{ k_i\},\{ k'_i\}} \phi_{N, \mathrm{in}}(\{k'_i\}),
\label{conv_phi}
\eeq
where $\phi_{N, \mathrm{in}}(\{k'_i\})$ and $\phi_{N, \mathrm{out}}(\{k_i\})$ are envelope functions of outgoing and incoming states depending on sets of outgoing $\{k_i\}$ and incoming $\{k'_i\}$ photonic energies, respectively, obeying the energy conservation $\sum_{i=1}^{N}k_i = \sum_{i=1}^N k'_i$. The convolution in \eq{conv_phi} is performed over the incoming set of momenta $d \{ k'_{i} \} = \frac{1}{N!} \prod_{i =1}^N d k'_{i}$.

A general diagrammatic approach to calculate the scattering matrix of a local emitter with an arbitrary level structure and transition amplitudes has been developed in Ref.~\cite{PG}. It gives results coinciding with a direct solution of the Lippmann-Schwinger equation \cite{shen-fan-LS}.

A theoretical study of the scattering off an array of distributed scatterers is more involved, since one has to take into account interference effects. For a single photon scattering, an evaluation of the scattering matrix is facilitated by an application of the transfer matrix method \cite{transM2,transM1}. For the scattering of two or more photons, there is no general prescription on how to compute the exact scattering matrix, and the complexity of this task is determined by the interplay of interference and correlation effects. There are few numerical results in the literature  \cite{baran1,baran2} on this problem.

The model \eq{model}, however, affords a considerable simplification based on the unidirectional propagation of light: No backscattering can happen during each scattering event. Moreover, all photons travel with the same group velocity. For this reason the interference does not occur, and the result of scattering does not depend on the travel time between emitters. Therefore, the scattering from one emitter happens independently of any other emitter, and the net result of the scattering  off an array of emitters is represented by the {\it convolution property} \cite{PG}
\beq
S^{(N)}_{M; \{ k_{i_M} \}, \{ k_{i_0}\} } ( \{ \Delta_a \}, \{ \kappa_a \}) &=& \int \left( \prod_{b=1}^{M-1} d \{ k_{i_b} \} \right)S^{(N)}_{\{ k_{i_M} \}, \{ k_{i_{M-1}}\} } ( \Delta_M ,  \kappa_M )  \nonumber\\
&\times&S^{(N)}_{\{ k_{i_{M-1}} \}, \{ k_{i_{M-2}}\} } (  \Delta_{M-1} ,  \kappa_{M-1} ) \ldots\nonumber \\
&  \times & S^{(N)}_{\{ k_{i_2} \}, \{ k_{i_1}\} } (  \Delta_2 ,  \kappa_2) S^{(N)}_{\{ k_{i_1} \}, \{ k_{i_0}\} } (  \Delta_1 ,  \kappa_1 ),
\label{convolution-property}
\eeq
where $ S^{(N)} (\Delta_a , \kappa_a)$ is the $N$-photon scattering matrix of the $a$-th emitter.

The property \eq{convolution-property} is very basic. We summarize the condition
under which it holds: 1) unidirectional nature of the spectrum; 2) a constant group velocity of the incoming wavepacket; and 3) linear and energy-independent coupling between the photons and the emitters. The independence of \eq{convolution-property} of emitters' positions $x_a$ lies at the origin of many {\it universal} properties of the outgoing photonic states which we discuss in the following.

A combination of methods to evaluate $S^{(N)}_{M=1}$ of a single emitter \cite{shen-fan-LS,PG} with the convolution property \eq{convolution-property} provides
a general platform for calculating scattering outcomes off arrays composed of emitters with an arbitrary complex structure of levels and transitions between them. For example, one can use three-level emitters with $S_{+}=g_{31}|3\rangle\langle 1|+g_{32}|3\rangle\langle 2|$ ($\Lambda$-scheme),
$S_{+}=g_{31}|3\rangle\langle 1|+g_{21}|2\rangle\langle 1|$ ($V$-scheme), or $S_{+}=g_{32}|3\rangle\langle
2|+g_{21}|2\rangle\langle 1|$ ($\Sigma$-scheme). One can even combine emitters of different types along the line of light's propagation. In all such cases the scattering matrix $S^{(N)}_{M}$ can be explicitly determined. Once we understand how its properties depend on the emitters' parameters, we obtain a powerful tool to {\it engineer} correlated multiphoton states.

In this paper, we concentrate on the description of the model \eq{model}. We also remark that its alternative solution in the case of identical couplings $\kappa_a = \kappa$ was obtained  by the means of the Bethe Ansatz \cite{RuYu84,Yu85,Yu88,YR}, which we use as a benchmark to verify our approach based on the usage of the convolution property \eq{convolution-property}.

\subsection{Single-photon scattering}

Let us now discuss an application of the general theory formulated above to the scattering of few-photon wavepackets  off an array of $M$ emitters. 

We start from the most basic case of single-photon scattering. In the following we will neglect losses setting $\kappa'=0$. This scattering is purely elastic: A photon scattered off a single emitter with parameters $\Delta_a$ and $\kappa_a$ just acquires an additional phase, which defines the scattering matrix
\beq
S_{kk'}^{(1)} = \delta_{kk'} \frac{k- \Delta_a - i \kappa_a/2}{k -\Delta_a +i \kappa_a/2} \equiv \delta_{kk'} e^{i \varphi_k^{(a)}}.
\label{eq:S1}
\eeq
Furthermore, phases acquired on every individual scatterer are additive, in accordance with the convolution property \eq{convolution-property}. This gives the single-photon scattering matrix of an array of $M$ emitters
\beq
S_{M; kk'}^{(1)} = \delta_{kk'} \prod_{a=1}^M \frac{k- \Delta_a - i \kappa_a/2}{k -\Delta_a +i \kappa_a/2} \equiv \delta_{kk'} e^{i \sum_{a=1}^M \varphi_k^{(a)}},
\label{eq:S1M_mom}
\eeq

The corresponding outgoing single-photon wavefunction at point $x$ and time $t$ can be decomposed as
\beq
\phi_{1,\mathrm{out}} (x-t)&\equiv &\int d x' S^{(1)}_M (x-t,x') \phi_{1,\mathrm{in}} (x') \nonumber\\
&=& \phi_{1,\mathrm{in}} (x-t) + \phi_{1,\mathrm{scatt}} (x-t),
\eeq
where $S^{(1)}_M (x,x')$ is the coordinate representation of \eq{eq:S1M_mom}, and $\phi_{1,\mathrm{scatt}}$ is the scattered part of the outgoing wavefunction.

To understand the structure of $\phi_{1,\mathrm{scatt}}$ we consider the limiting case of  identical emitters $\Delta_a \to \Delta$, $\kappa_a \to \kappa$, and obtain the expression (see the Appendix for details)
\beq
\phi_{1,\mathrm{scatt}} (x-t) = - \int d x'  \Theta (x') \kappa L^{(1)}_{M-1} (\kappa x') e^{- (i \Delta +\kappa /2) x'} \phi_{1, \mathrm{in}} (x' +x -t),
\label{eq:1phot_equal}
\eeq
where $L_{M-1}^{(1)} (x)$ is the associated Laguerre polynomial. An appearance of the polynomial behavior is remarkable, and we next discuss its implications.

For a realistic scattering experiment we specify the initial wavepacket $\phi_{1, \mathrm{in}}(x)= \frac{1}{\sqrt{\sigma \sqrt{\pi}}} e^{i (\Delta+\delta) x - \frac{x^2}{2\sigma^2}}$. In the momentum space it corresponds to a Gaussian distribution around $k' = \Delta +\delta$ with the variance $1/\sigma$, where $\delta$ is the detuning. Assuming that $\delta^{-1}, \kappa^{-1} \gg \sigma$, we obtain
\beq
\phi_{1,\mathrm{scatt}} (x-t) \approx -  \sqrt{2 \sigma \sqrt{\pi}}  \Theta (t-x) \kappa L^{(1)}_{M-1} (\kappa (t-x)) e^{-\kappa (t-x)/2} e^{i \Delta (x-t)} .
\label{eq:1photonScatt}
\eeq

\begin{figure}[h]
    \includegraphics[width=14cm]{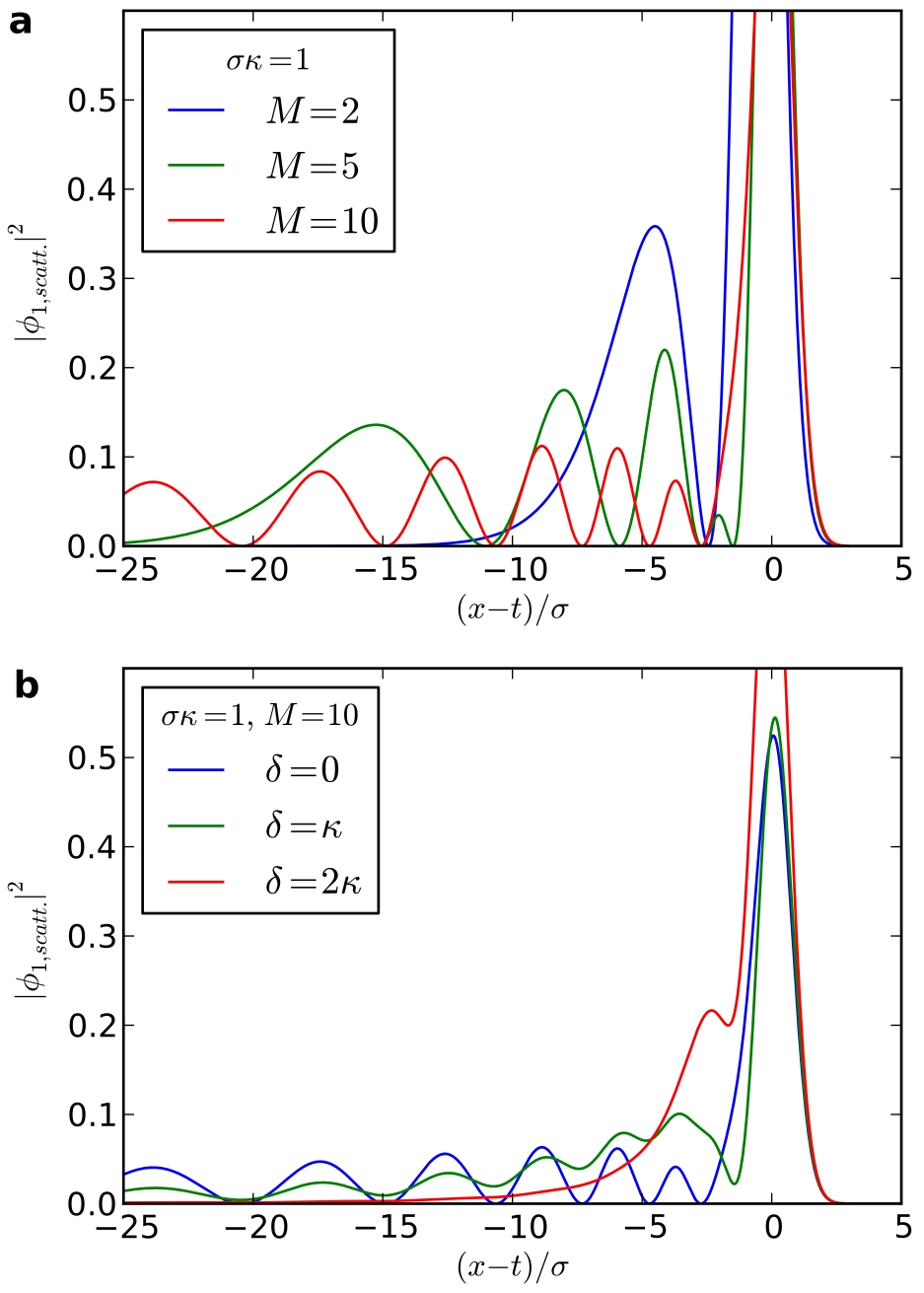}
\caption{{\bf One-photon scattering off $M$ emitters.} {\bf a,} One-photon scattering of the Gaussian wavepacket (variance $\sigma$) off $M$
 two-level emitters for detuning~$\delta=0$. 
 The oscillatory structure of the outgoing wavepacket is described by
 Eq.~\eq{eq:1photonScatt}.
 {\bf b,} One photon scattering  off $M=10$  two-level emitters for various values of
 detuning~$\delta$. The initial state is the same as in~{\bf a}. At large detunings the oscillations are suppressed.}
    \label{fig:1photonVsMSigma}
\end{figure}

In Fig.~(\ref{fig:1photonVsMSigma}) we plot~$|\phi_{1,\mathrm{scatt}}|^2$ for various numbers of emitters $M$ and detunings $\delta$ \cite{mask}.
These results clearly manifest the  universal character of scattering
in the system under consideration: The dependence of the outgoing wavepacket on the positions of emitters is absent, while the minima are determined by universal numbers --  the zeroes of the Laguerre polynomials. Thus, if the position of the first node is known, the subsequent nodes can be
determined from  $L^{(1)}_{M-1}(x)$. An emergence of nodes is accounted by time delays on each emitter, which eventually leads to the fragmentation of the incoming wavepacket into $M-1$ pieces. 

\subsection{Two-photon scattering}
 
Let us next consider  two-photon scattering. It can happen in two different ways: 1) via the {\it elastic} scattering of two individual photons; 2) via the {\it inelastic} scattering of two photons exchanging energy with each other. The existence of the second possibility is characteristic for {\it interacting} systems, which leads to an emergence of correlated states of photons. The corresponding two-photon scattering matrix is represented by a sum of reducible (elastic) and irreducible (inelastic) terms
\beq
S_{k_1 k_2, k'_1 k'_2}^{(2)} = S_{k_1 k'_1}^{(1)} S_{k_2 k'_2}^{(1)} +S_{k_1 k'_2}^{(1)} S_{k_2 k'_1}^{(1)} + i \mathcal{T}_{k_1 k_2 , k'_1 k'_2}^{(2)} .
\label{S2_dec}
\eeq
This picture of scattering mechanisms is important for the interpretation of scattering results below.  

The amplitude of  inelastic scattering off a single emitter $M=1$ is known
\cite{shen-fan-LS,PG}
\beq
 \mathcal{T}_{k_1 k_2 , k'_1 k'_2}^{(2)}  = \frac{\kappa^2}{\pi}  \frac{K - 2 \Delta + i \kappa}{\prod_{p}(p - \Delta +i \kappa/2) } \delta_{k_1 +k_2, k'_1 +k'_2},
\label{eq:T1}
\eeq
where $K= k_1+k_2 =k'_1 +k'_2$ and where $p\equiv(k_{1},k_{2},k_{1}',k_{2}')$. On the basis of this expression one can explain the bunching property of photons in one-dimensional waveguides. Most clearly this can be viewed in the coordinate representation, see below. 

In order to find an explicit expression of $S_{M; k_1 k_2, k'_1 k'_2}^{(2)}$ for arbitrary $M$ it is necessary to compute $M-1$ convolutions in \eq{convolution-property}, each being represented by a twofold momentum integration. As both $S^{(1)}_{kk'}$ in \eq{eq:S1} and $\mathcal{T}_{k_1 k_2 , k'_1 k'_2}^{(2)}$ in \eq{eq:T1} have a simple pole structure, this becomes a routine task.

A direct pathway to $S_M^{(2)}$ is available in the case when the couplings of the field to all emitters are identical, $\kappa_a = \kappa$. The two-photon (and also $N$-photon) scattering matrix in the coordinate representation can be then determined by the Bethe Ansatz method, and it reads \cite{Yu88}
\beq
S_M^{(2)} (y_1, y_2 , z_1 ,z_2) &=& \frac{1}{(2 \pi)^2} \int d k_1 dk_2 d k'_1 d k'_2 e^{i k_1 y_1 +i k_2 y_2 - i k'_1 z_1 - i k'_2 z_2} S_{M; k_1 k_2, k'_1 k'_2}^{(2)} \nonumber \\
&=& \Theta (z_1 - z_2)\int_{\gamma_1} \frac{d \lambda_1}{2 \pi} \int_{\gamma_2} \frac{d \lambda_2}{2 \pi} e^{i \lambda_1 (y_1 - z_1) + i \lambda_2 (y_2 - z_2)} \nonumber\\
&\times&\left[ 1 - \frac{2 i \kappa \Theta (y_2 - y_1)}{\lambda_1 - \lambda_2 + i \kappa}\right] \nonumber \\
&  \times &\left( \prod_{a=1}^M \frac{\lambda_1 - \Delta_a - i \kappa/2}{\lambda_1 - \Delta_a + i \kappa/2}\right) \left( \prod_{b=1}^M \frac{\lambda_2 - \Delta_b - i \kappa/2}{\lambda_2 - \Delta_b + i \kappa/2} \right) \nonumber\\ 
&+& (y_1 \leftrightarrow y_2) \cdot (z_1 \leftrightarrow z_2) .
\label{SM}
\eeq
Here the contours of integration $\gamma_1$ and $\gamma_2$ are chosen in such a way that $\lambda_1 = r_1 - i \kappa/2 + i \epsilon$, $\lambda_2 = r_2 +i \kappa/2 +2 i \epsilon$, $r_{1,2} \in \mathbb{R}$, and $\epsilon >0$ is an infinitesimal parameter. The expression \eq{SM} is much more compact than \eq{convolution-property}, since it contains a twofold integration instead of a $2^{M-1}$-fold one. Nevertheless, it is still useful to use both approaches of Refs.~\cite{PG} and \cite{Yu88} to unravel all properties of the scattering matrix $S_M^{(2)}$. 

One can immediately note the fundamental properties of the two-photon scattering matrix \eq{SM}. First, it does not contain any dependence on the positions of emitters, therefore the scattering is robust with respect to variations of the latter. Second, the expression \eq{SM} is invariant under a permutation of emitters -- the products over the set of emitter do not change. Therefore, the scattering results do not depend on the ordering of emitters. They rather appear to be characteristic of sets of emitters than of individual emitters. This feature of \eq{SM} allows us to re-order emitters for the purposes of computational efficiency, in particular in the expression \eq{convolution-property}.

Modeling the incoming state by the Gaussian two-photon wavepacket
\beq
    \phi_{2,\mathrm{in}} (x_1,x_2) &=& \frac{1}{\sqrt{2\sigma\mu\pi}}\exp[ i (\Delta+\delta) (x_1+x_2)]\\
&\times&\exp[- (x_1+x_2)^2/8\mu^2 - (x_1-x_2)^2/2\sigma^2 ], \nonumber
\label{eq:TwoPhotonIncomingWF}
\eeq
where $\mu$ and $\sigma$ are the variances of the center-of-mass coordinate $\frac{x_1 +x_2}{2}$ and the relative coordinate  $d=x_1 -x_2$ distributions, respectively, we are mainly interested in the regime $\mu \gg \sigma$. In the limit
$\mu\rightarrow\infty$, the total energy $K=k_1+k_2=k'_1+k'_2$ of the two photons is approximately conserved at the value $K=2 \Delta + 2 \delta$, and the incoming wavepacket \eq{eq:TwoPhotonIncomingWF} acquires the factorized form
\beq
 \phi_{2,\mathrm{in}} (K ,d) &=& \frac{1}{\sqrt{2 \pi}} \int d X e^{-i K X} \phi_{2,\mathrm{in}} \left(X +\frac{d}{2},X -\frac{d}{2} \right) \nonumber\\
 &\approx &\frac{\sqrt{\mu}}{\sqrt{2 \sqrt{\pi} }} e^{-\frac{\mu^2}{2} (K- 2 \Delta - 2 \delta)^2} \phi_{2,\mathrm{in}} (d) ,
\label{fact}
\eeq
where $\phi_{2,\mathrm{in}} (d) = \frac{1}{\sqrt{\sigma \sqrt{\pi}}} e^{-\frac{d^2}{2 \sigma^2}}$. 

In this setting, the entire effect of scattering is visible in the relative part $\phi_2 (d)$ of the two-photon wavefunction: If the energy is conserved, then  $\phi_{2,\mathrm{out}} (K ,d)$ can be also factorized like \eq{fact}, and we can
express the relative part  $\phi_{2, \mathrm{out}} (d)$ of $\phi_{2, \mathrm{out}} (K,d)$ through $\phi_{2, \mathrm{out}} (d)$
\beq
\phi_{2, \mathrm{out}} (d) &=& \int d (d') S_M^{(2)} (d,d') \phi_{2, \mathrm{out}} (d')
\label{phi_out}
\eeq
by means of  the {\it relative} scattering matrix $S_M^{(2)} (d,d')$ depending on relative coordinates of photons $d$ and $d'$ in final and initial states, respectively. We also note that $S_M^{(2)} (d,d')$ also depend parametrically on $\delta$ which measures a detuning of the total energy from the two-photon resonance $2 \Delta$. 

We note that the representation of a scattering wavefunction in relative coodinates of photons is very advantageous, because of its direct relation to the correlation function $G^{(2)} (\tau)= \langle \mathrm{out} | a^{\dagger} (x) a^{\dagger} (x +\tau) a (x +\tau)  a(x) | \mathrm{out}\rangle = 4 | \phi_{2, \mathrm{out}} (d = \tau)|^2$, which is an important measure of correlation effects between photons.

In the case of identical emitters $\Delta_a = \Delta$, $\kappa_a =\kappa$, the relative scattering matrix $S_M^{(2)} (d,d')$ is explicitly given by
\beq
S_M^{(2)} (d,d') &=& \delta (|d| - |d'|) \label{Sdelta} \\
&-& \frac{i}{(M-1)!} \frac{\partial^{M-1}}{\partial s^{M-1}} \left\{ \left[ e^{i \left| |d| - |d'|\right| (\delta+i \kappa/2 -s)} +e^{i ( |d| + |d'|) (\delta+i \kappa/2 -s)} \right]\right. \nonumber\\
&\times& \left.\frac{(s-i \kappa)^M (s - 2 \delta)^M}{(s- 2 \delta - i \kappa)^M}  \right\}_{s=0} \label{Selast} \\
&-& \frac{ \kappa }{(M-1)!} \frac{\partial^{M-1}}{\partial s^{M-1}} \left\{ e^{i (|d|+|d'|) (\delta +i \kappa/2 -s) } \frac{(s-i \kappa)^M}{ (s - \delta)} \right.\nonumber\\
&\times&\left. \left[ \frac{(s - 2 \delta)^M}{(s - 2 \delta - i \kappa)^M} - \frac{s^M}{(s + i \kappa)^M}\right] \right\}_{s=0}, \label{Sinelast}
\eeq
see the Appendix for details of the derivation. This expression contains interesting effects which we discuss below.

Let us first remark the basic properties of  $S_M^{(2)} (d,d')$. First, we note that it is {\it symmetric},
\beq
S_M^{(2)} (d,d') = S_M^{(2)} (d',d).
\label{symmetry_red}
\eeq
Second, it obeys the {\it reality} condition
\beq
S_M^{(2)*} (d, d'; \delta) = S_M^{(2)} (d, d'; - \delta).
\label{reality_red}
\eeq
Third, the convolution property is also fulfilled in the relative coordinates
\beq
S_M^{(2)} (d,d') &=& \int_0^{\infty} \prod_{b=1}^{M-1}  d (d_b) S^{(2)} (d,d_{M-1}) S^{(2)} (d_{M-1}, d_{M-2})\times \nonumber\\&\ldots &\times S^{(2)} (d_2 , d_1) S^{(2)} (d_1 , d'). 
\label{convolution_red}
\eeq
Fourth, the unitarity condition implies that
\beq
\delta (|d| - |d'|) = \int_0^{\infty} d (d_1) S_M^{(2)*} (d, d_1) S_M
^{(2)} (d_1 ,d').
\label{unitarity_red}
\eeq

The listed properties impose rigid constraints on possible scattering outcomes. In particular, in the resonance case $\delta =0$ (which in the present context means only that the total energy $K$ mtaches with $2 \Delta$, while the energies of both photons can differ from each other) the scattering result does not depend on the number $M$, but rather on its parity
\beq
S_{M \,\, \mathrm{odd}}^{(2)} (d,d') &=& S_{M=1}^{(2)} (d,d') , \label{oddS} \\ 
S_{M \,\, \mathrm{even}}^{(2)} (d,d')&=& S_{M=2}^{(2)} (d,d') = \delta (|d| - |d'|). \label{evenS}
\eeq
Moreover, the latter case of an array with an even number of scatterers is {\it transparent} for incident light,
\beq
\phi_{2, \mathrm{out}}^{M \,\, \mathrm{even}} (d) = \phi_{2, \mathrm{in}} (d) .
\label{eq:TwoPhotonsOutEven}
\eeq
To substantiate these conclusions, it is sufficient to notice that the unitarity condition \eq{unitarity_red} for $M=1$ in combination with the symmetry \eq{symmetry_red} and reality \eq{reality_red} conditions yields $S_{M=2}^{(2)} = \delta (|d|-|d'|)$, which holds by virtue of \eq{convolution_red}. The generalization for arbitrary $M$ follows from the further application of the convolution property \eq{convolution_red}. 

The same result \eq{oddS}, \eq{evenS} follows form the explicit expression for $S_{M=1}^{(2)} (d,d')$ given by \eq{Sdelta}-\eq{Sinelast}. This consideration  elucidates the crucial role of the inelastic processes for the emergence of the parity effect \eq{oddS},\eq{evenS}: The elastic contribution \eq{Selast} vanishes at $\delta =0$, and only the inelastic contribution \eq{Sinelast} makes the scattering off an odd number of emitters and off an even number of emitters distinguishable. In addition, we note that the parity effects in a setup similar to ours and driven by a {\it classical} field have been discussed in Ref.~\cite{stannigel}.

The transparency effect expressed in \eq{evenS} can be straightforwardly generalized to arrays of emitters with a set of different detunings $\{ \delta_a \}$ obeying a constraint such that for each $\delta_a$ there exists $\delta_{\bar{a}} = -\delta_a$ (for even $M$). In fact, reshuffling the scattering matrices of individual scatterers in \eq{convolution_red} (which is allowed by the permutation symmetry discussed above), we can arrange them pairwise. The convolution within each pair $(\delta_a , \delta_{\bar{a}})$ yields the identity ($\delta$-function) by virtue of the symmetry, the reality, and the unitarity conditions; a convolution of the identities is again the identity.

Since the $M=1$ case is important for the understanding of the scattering off an odd number of emitters, we briefly revisit it. For $\delta =0$ we have
\beq
S^{(2)} (d,d') = \delta (|d| - |d'|) -  2  \kappa e^{- \kappa (|d|+|d'|)/2 }.
\eeq
For the Gaussian initial condition we obtain
\beq
    \phi_{2, \mathrm{out}} (d) &=&
   \frac{1}{\sqrt{\sigma \sqrt{\pi}}} 
    \left[ 
    e^{-\frac{d^2}{2\sigma^2}} - \sqrt{2\pi} \kappa\sigma
    e^{-\frac{\kappa |d|}{2}+\frac{\sigma^2\kappa^2}{8}}
    \mbox{erfc}\left( \frac{\sigma\kappa}{2\sqrt{2}} \right)
    \right]\nonumber\\ 
    & \stackrel{\sigma \gg \kappa^{-1}, |d|}{\approx}&  \frac{1}{\sqrt{\sigma \sqrt{\pi}}} 
    \left[ 1- 4  e^{-\frac{\kappa |d|}{2}} \right].
\label{eq:TwoPhotonsOutOdd}
\eeq
This wavefunction describes a bound state of two photons. If the initial variance $\sigma$ of  relative distances between photons is sufficiently large, $\sigma \gg \kappa^{-1}$, then the final distance is distributed on the smaller length scale $\kappa^{-1}$, which indicates an emerging effective attraction between the photons.

\begin{figure}[h]
   \includegraphics[width=14cm]{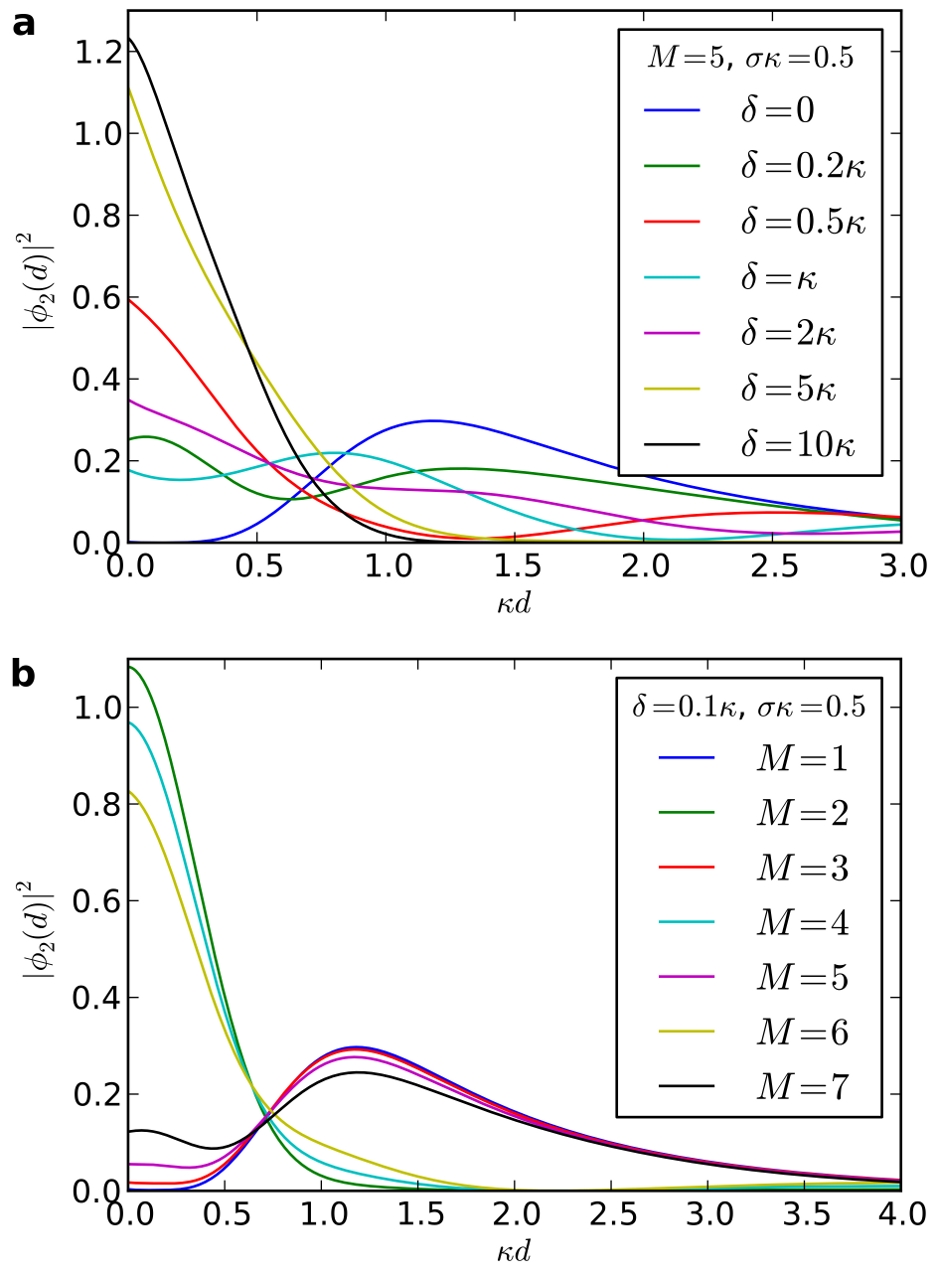}
   \caption{ {\bf Two-photon scattering off $M$-atoms.} 
    {\bf a,}
    Scattering of the Gaussian two-photon wavepacket, Eq.~\eq{eq:TwoPhotonIncomingWF}, off $M=5$ two-level emitters as a function of the relative coordinate~$d=x_1-x_2$. The results change qualitatively from antibunching at small $\delta$ to bunching at large $\delta$.
    {\bf b,}
    The same dependence as in {\bf a,} for fixed $\delta$ and various numbers of emitters $M$. The result exhibits the clearly pronounced parity effect of Eqs.~\eq{oddS},\eq{evenS}.}
    \label{fig:2photonVsSigma}
\end{figure}

Another important consequence for the scattering off an odd number of emitters which can be derived from \eq{eq:TwoPhotonsOutOdd} is that for a special choice of the parameter $\sigma \kappa$ one can observe the antibunching behavior of photons. It is characterized by the vanishing value of  $G^{(2)} (0) \propto |\phi_{2, \mathrm{out}} (0)|^2$, which indeed happens at $\sigma \kappa \approx 0.5$. In contrast, for an even number of emitters  $G^{(2)} (0)$ does not vanish for any $\sigma \kappa$, which indicates a tendency towards bunching. Therefore we use the value $\sigma \kappa =0.5$ in Fig. (\ref{fig:2photonVsSigma}b) depicting $|\phi_{2, \mathrm{out}} (d)|^2$  in order to emphasize the qualitative difference between scattering off even and odd numbers of emitters.

A much richer structure of an outgoing wavefunction appears at a finite detuning $\delta$. Assuming now $\delta^{-1}, \kappa^{-1} \gg \sigma$, we observe that it develops a polynomial dependence on $d$
\beq 
 \phi_{2, \mathrm{out}} (d) =  \phi_{2, \mathrm{in}} (d) -   \sqrt{2 \sigma \sqrt{ \pi}} \kappa  e^{- |d| (\kappa/2 - i \delta)} P_{M-1} (\kappa |d|; \delta/\kappa) ,
\eeq
where the polynomial of the order $M-1$
\beq
P_{M-1} (x; \hat{\delta}) =
 \frac{e^x}{(M-1)!} \frac{\partial^{M-1}}{\partial s^{M-1}} \left\{  e^{-  s x } s^M  \frac{ ( s - i \hat{\delta} - \frac12) (s  - 2 i \hat{\delta}-1)^M}{ (s  - i \hat{\delta}-1) (s  - 2 i \hat{\delta} )^M}  \right\}_{s=1}  
\eeq
has complex-valued coefficients. 

At a large detuning $\delta \gg \kappa$ the elastic scattering dominates, $P_{M-1} (x; \hat{\delta})$ turns into the Laguerre polynomial $L_{M-1}^{(1)} (x)$ with the real-valued coefficients, and $\phi_{2, \mathrm{out}} (d)$ exhibits the same behavior as the one-photon scattering function Eq.~\eq{eq:1phot_equal}: The probability $|\phi_{2, \mathrm{out}} (d)|^2$ oscillates on the scale $\kappa^{-1}$ featuring precisely $M-1$ nodes. We note that the real-valuedness of the coefficients of $P_{M-1} (x; \hat{\delta})$ is important for the presence of the nodes in $|\phi_{2, \mathrm{out}} (d)|^2$, which are destroyed at a small detuning $\delta \ll \kappa$ by the emerging effective interaction between photons.

Thus, tuning  $\delta$ from $\delta \gg \kappa$ to $\delta \ll \kappa$ we observe a qualitative change in the scattering properties (see Fig.~\ref{fig:2photonVsSigma}a), characterized by the increasing role of the inelastic processes. The formation of bound states, transitions from bunching to antibunching, and the clearly pronounced parity effect give full evidence of strong correlations between photons resulting from their scattering off arrays of emitters in a one-dimensional chiral channel.

\subsection{Robustness of correlated states}

\begin{figure}[h]
    \includegraphics[width=12.5cm]{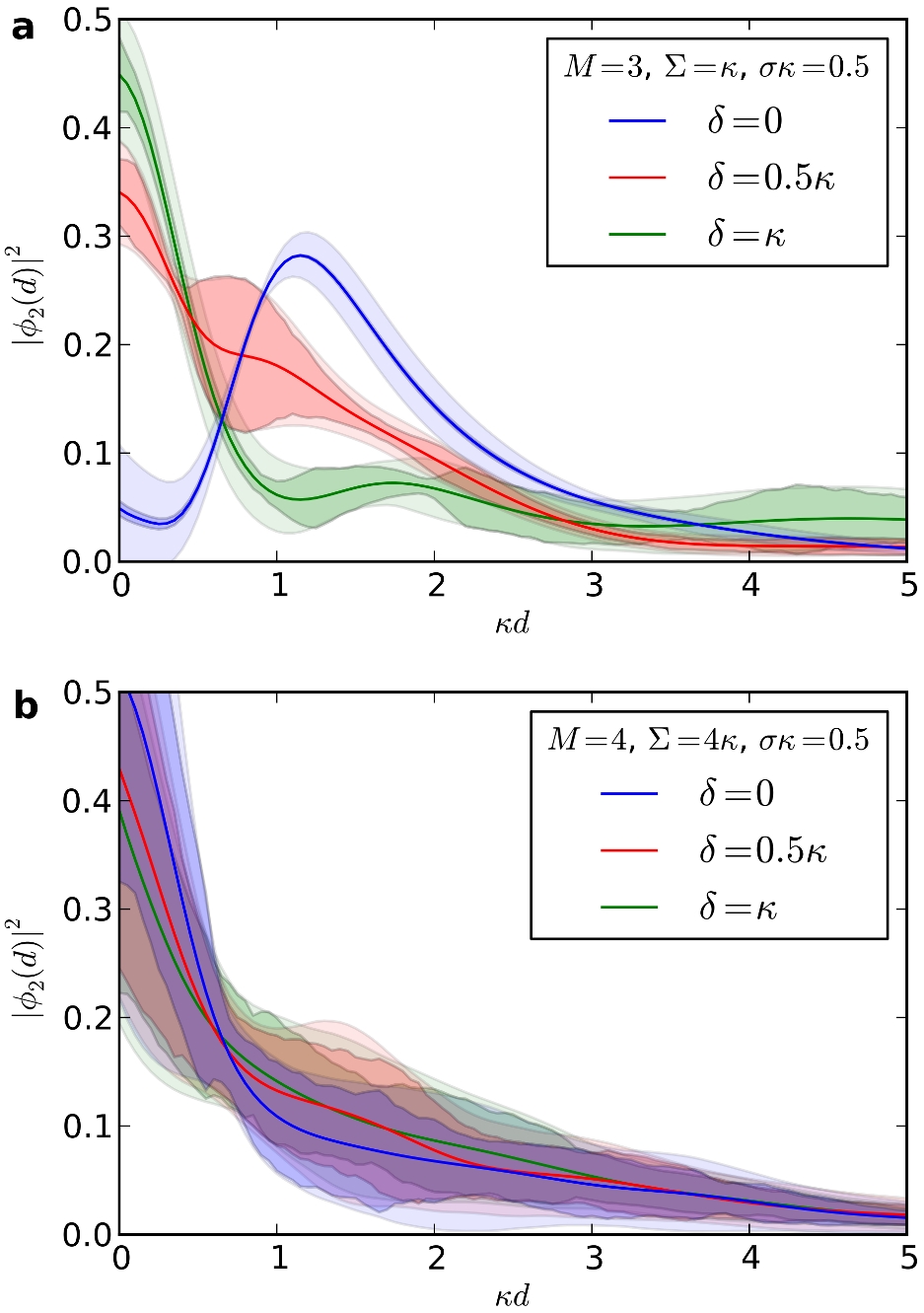}
  \caption{ {\bf Two-photon scattering off $M$-atoms with disorder.}
  {\bf a,} The scattering of a Gaussian two-photon wavepacket off $M=3$ two-level emitters averaged over the fluctuating transition frequencies. The latter are modeled by independent random variables which are normally distributed around the mean value $\delta$ with the variance $\Sigma$. 
  The solid lines correspond to the mean values of the distribution Eq.~\eq{prob_distrib}. 
  The filled regions denote the median (darker) and mean (lighter) absolute deviation.  
        {\bf b,} The same dependence as in {\bf b}, but for $M=4$ emitters. Both figures demonstrate that the qualitative properties of scattering states are robust against the fluctuations in the transition frequencies.
        }
    \label{fig:disorder}
\end{figure}

It has been already discussed above that scattering results are independent of the positions of emitters, which ensures their robustness against fluctuations of the latter. Next, we study how robust scattering results are with respect to randomness in the emitters' transition frequencies. To this end, we consider detunings $\delta_a$ as random variables which are normally distributed around the mean value $\delta$ with the variance $\Sigma$, and evaluate the probability distribution 
\beq
& &P [|\phi_{2, \mathrm{out}} (d)|^2] \nonumber\\
&=& \int \prod_{a'=1}^M   \left( \frac{e^{- \frac{(\delta_{a'} - \delta)^2}{2 \Sigma^2}} d \delta_{a'}}{\Sigma \sqrt{2 \pi}} \right)  \delta \left(|\phi_{2, \mathrm{out}} (d)|^2 - |\phi_{2, \mathrm{out}} (d; \{\delta_a \})|^2 \right)
\label{prob_distrib}
\eeq
with the help of the exact expression for $|\phi_{2, \mathrm{out}} (d; \{\delta_a \})|^2$ based on Eqs.~\eq{SM}-\eq{Ca}. In Figs.~(\ref{fig:disorder}) we plot typical results of this averaging showing mean values as well as median and mean absolute deviations of the distribution \eq{prob_distrib}. We observe that for not so large $\Sigma \lesssim \kappa$, this distribution is sufficiently narrow so that it does not mask the qualitative effects (parity, antibunching, etc.) discussed above. Thus, we conclude that our results are robust with respect to the fluctuations in transition frequencies as well.

\subsection{Effect of losses}

To estimate an effect of losses due to the spontaneous emission out of the one-dimensional modes we can again use the formulas \eq{Sdelta}-\eq{Sinelast}, replacing $\delta \to \delta +i \kappa'/2$, where $\kappa'$ is the spontaneous emission rate (considered to be small $\kappa' \ll \kappa$). After this replacement the scattering matrix is no longer unitary, which means that the norm of photonic wavefunctions is not preserved. To what extent does this modify the parity effect? By an explicit calculation for $\delta=0$ one can find that $S^{(2)}_{M=1} = \delta (|d|-|d'|) - \frac{\kappa (2 \kappa+\kappa')}{\kappa +\kappa'} e^{- (\kappa +\kappa') (|d| +|d'|)/2} - \frac{\kappa \kappa'}{\kappa +\kappa'} e^{-(\kappa +\kappa') ||d|-|d'||/2}$ and $S^{(2)}_{M=2}= \delta (|d| - |d'|) -\kappa' (6 - 2 \kappa (d+d')) e^{- (\kappa +\kappa') (|d| +|d'|)/2} - 2 \kappa' e^{- (\kappa +\kappa') ||d| -|d'||/2} + O (\kappa'^2/\kappa)$. The terms $\sim \kappa'$ in the latter expression estimate (when divided by $\kappa$): a) which part the norm of an initial two-photon wavepacket leaks out of the one-dimensional channel; b) a deviation from the full restoration of the initial wavepacket shape after scattering on two emitters. On this basis we conclude that the deviation $| \phi_{2, \mathrm{out}}^{M \, \mathrm{odd}} (d)- \phi_{2, \mathrm{out}}^{M=1} (d)|$ from the perfect parity effect is accumulated on each scatterer, and thereby it is proportional to $M \frac{\kappa'}{\kappa}$.

\section{Discussion}

We have suggested a novel approach to producing strongly correlated states of two and more photons in a highly
controllable way. Our setup uses the edge states of photonic topological insulators with multi-level emitters coupled to it. The scattering in this setup possesses the following universal features.

First, multi-particle scattering does not depend on the positions of emitters. This is a general consequence of the absence of backscattering, and it makes the scattering outcomes robust against the fluctuations in the emitters' coordinates. This property also provides us with a powerful theoretical tool for calculating the scattering results based on the convolution property \eq{convolution-property}.

Second, the scattered wavepacket has a polynomial structure and the minima of the outgoing pulse are given by the
zeroes of this polynomial. In particular, single-particle scattering (and the reducible part of multi-particle scattering in general) is described by Laguerre polynomials. In single-photon scattering, the outcome looks like a wavepacket fragmented between the nodes -- the zeroes of this polynomial. The emergence of the nodes is accounted by time delays on each emitter. This leads to the non-monotonic behavior of the $g_{2}$ function thus signifying correlation between photons. In multi-photon scattering, this picture takes place at large detunings $\delta$, i.e., in the regime where the elastic processes dominate over the inelastic ones.

Third, the role of the inelastic processes enhances as we tune to small values of $\delta$. These processes lead to the emergence of the effective interaction between photons, which lies at the origin of the exciting effects: The parity-dependent scattering and the antibunching of photons. In particular, we have observed for the first time that at a vanishing detuning the two-photon scattering matrix in the chiral edge channel only depends on the parity of the number of emitters, and does not depend on the number itself. We substantiated our observation by the fundamental symmetry arguments. For an even number of emitters, we proposed certain array configurations  which are transparent to arbitrary two-photon wavepackets. For an odd number of emitters, we discussed how to adjust the parameters of the initial two-phonon wavepackets in order to observe the antibunching behavior -- the effective repulsion of photons -- in the one-dimensional geometry.

The physical picture behind the parity effect is the following. After the first incident photon is absorbed by the first emitter, the latter becomes transparent for the second photon. Thus, the first photon is delayed by the time  $\sim \kappa^{-1}$ with respect to the second photon. This leads to antibunching of photons via scattering on a single emitter. If the second emitter is present, it absorbs the second photon and captures it during the time $\sim \kappa^{-1}$. This gives the first photon a possibility to compensate the lag, and both photons end up with the same relative distance as it was in the incoming state (note that the compensation is exact, only if the resonance condition is fulfilled, and if the emitters are identical). Repeating these arguments for odd and even numbers of identical emitters, we observe that the net result is the same as for one and two emitters, respectively. Deviations from the perfect conditions would lead to a difference between scattering results on, say, $M=1$ and $M=3$ emitters, but they still remain closer and qualitatively similar to each other, than to a scattering result on $M=2$ emitters. 

These exciting effects result from the combination of the reduced dimensionality, which increases the role of the photonic correlations, and the topological origin of the chiral edge states propagating without the backscattering. For their experimental observation, we specify below the necessary physical parameters.  Photonic correlations are a vital ingredient for building up working schemes of future photonic quantum technology \cite{PhotQTech}. We believe that the effects described here will be used in future functional quantum photonic devices or for quantum simulation of condensed matter systems using photonic setups.

{\it Physical realization}. To observe the quantum many-body effects of photons described above one needs
a photonic topological insulator analogous to the one studied in \cite{MIT,top-light-exp}. The
role of emitters can be played by either quantum dots or superconducting qubits, like, e.g., in
\cite{1-photon-GHz} and \cite{astafiev}. Available single-photon emitters made of quantum dots  and coupled to photonic crystal waveguides \cite{Yao,hoi,1-photon-rev} can also serve as a realization of our model. Typical bandgap frequencies of existing photonic topological insulators lying in the GHz range  match with the transition frequencies of single-photon emitters, which are necessary  to inject photons into the one-dimensional channel. The parameters provided in Ref.~\cite{MIT} allow us to estimate the topological photonic band gap $B\sim$0.27 GHz as well as the group velocity  $c/v_{g}\approx 175$.  The electromagnetic field is bounded within an area of $\sim$1600 mm$^2$, which gives $A_{0}/A_{eff}\sim 1$. The dielectric host material has a relatively large dielectric constant (see the Method summary section of Ref.~\cite{MIT}). On the basis of \eq{purcell} we conclude that the Purcell factor of the setup of Ref.~\cite{MIT} is of the order of a few tens. The other setup described in \cite{floquet} is made of an array of helix-like waveguides with the radius $R$. The spectral gap and the group velocity of the edge state (see Fig. 2c of Ref.~\cite{floquet}) depend nonmonotonically on $R$; the group velocity can even go down close to zero at $R\approx 17\mu$m, rendering $c/v_{g}$ very large. This leads us to much more optimistic values of $c/v_{g}\sim 10^{3}$ and a Purcell factor of the same order. A similar analysis of the third setup presented in \cite{haf1} yields a Purcell factor $\sim$20. A further enhancement of the Purcell factor in the near future can facilitate the experimental realization of our proposal.

In this paper we discussed the scattering results in arrays of two-level emitters. The present consideration, however, can be generalized to arrays of three- and four-level emitters, which are interesting due to  the experimentally observed effects \cite{abdumalikov,EIT-mucke,EIT-kampschulte} of the single-emitter electromagnetically induced transparency (EIT)  and the transistor-like behavior, and the recent proposals to implement four-level emitters in the engineering of quantum gates, e.g. CNOT \cite{baranger}. By combining emitters of different level structures one gains flexibility and the ability to
build optical schemes with desired correlations of the outgoing photonic wavefunctions. For an explicit evaluation of the outgoing state one can use the scattering formalism of Ref.~\cite{PG} in combination with the convolution property \eq{convolution-property}.

Currently available topological photonic crystals operate in the  GHz frequency range. This is related to the
difficulties in creating a large magneto-optical response with existing materials, which is needed for a nontrivial 
topology of the band structure. However, some recent developments \cite{magnetoplasmon}
in material science might lead to a significant enhancement of the magneto-optical response in the optical domain, 
which can be used to engineer topological photonic crystals in the optical range of frequencies.

Finally, we note that interacting photons in two dimensions can lead to anyonic statistics and fractional Hall
states \cite{2D}. The recent suggestion of realizing Dirac cone structure \cite{dirac} may pave the way
to exotic physics of interacting two-dimensional photonic fluids. Moreover, the recent classification of
interacting {\it bosonic} topological insulators \cite{wen} suggests a possible route to engineer other exotic states of light.

We expect that the novel correlated states of photons and other propagating bosons  will find applications in quantum information technology, optomechanics, and precision measurements.

\section*{Appendix: Technical details}
In the case of two-level emitters and identical couplings the model \eq{model} is exactly solvable by the Bethe Ansatz \cite{Yu88}. For a more general problem involving  arrays of emitters with an arbitrary level structure, a complementary diagrammatic approach to scattering is available in Ref.~\cite{PG}. 

Below we provide the derivation of formulas appearing in the main text based on these approaches.

\subsection*{Single-photon scattering}

First, we rewrite the single-photon scattering matrix \eq{eq:S1M_mom} in the coordinate representation
\beq
S^{(1)}_M (y,z) &=& \frac{1}{2 \pi} \int d k  dk' e^{i k y - i k' z} S_{M;kk'}^{(1)} \nonumber \\
&=& \delta (y-z) - \Theta (z-y) \sum_{a=1}^M \kappa_a \prod_{a'=1; a \neq a}^M \frac{\Delta_a - i \kappa_a/2 -\Delta_{a'} - i \kappa_{a'}/2}{\Delta_a - i \kappa_a/2 -\Delta_{a'} + i \kappa_{a'}/2}\nonumber\\
&\times& e^{- (i \Delta_a +\kappa_a /2) (z-y)}.
\label{eq:S1M}
\eeq
The corresponding scattered part $\phi_{1,\mathrm{scatt}}$ is given by
\beq
 \phi_{1,\mathrm{scatt}} (x-t) &=& - \int d x'  \Theta (x') \sum_{a=1}^M \kappa_a \prod_{a'=1; a' \neq a}^M \frac{\Delta_a - i \kappa_a/2 -\Delta_{a'} - i \kappa_{a'}/2}{\Delta_a - i \kappa_a/2 -\Delta_{a'} + i \kappa_{a'}/2}\nonumber\\
 &\times& e^{- (i \Delta_a +\kappa_a /2) x'} \phi_{1, \mathrm{in}} (x' +x -t).
\eeq

In the limit $\Delta_a \to \Delta$, $\kappa \to \kappa_a$  we obtain
\beq
S_M^{(1)} (y,z) &=&
\frac{1}{2 \pi} \int d k e^{i k (y-z)} \left( 1 - \frac{ i \kappa}{k -\Delta + i \kappa/2} \right)^M \\
&=& \delta (y-z) + \sum_{m=1}^M C^M_m \frac{(- i \kappa)^m}{2 \pi} \int d k \frac{e^{i k (y-z)}}{(k - \Delta + i \kappa/2)^m} \nonumber \\
&=&  \delta (y-z) - i \Theta (z-y) \sum_{m=1}^M C_m^M \frac{(- i \kappa)^m [i (y-z)]^{m-1}}{(m-1)!} e^{(i \Delta +\kappa/2) (y-z)} \nonumber \\
&=& \delta (y-z) - \kappa \Theta (z-y) L_{M-1}^{(1)} (\kappa (z-y))   e^{(i \Delta +\kappa/2) (y-z)} ,
\label{ftss}
\eeq
where $L_{M-1}^{(1)} (x)$ are  the associated Laguerre polynomials.

\subsection*{Two-photon scattering}

The two-photon scattering matrix \eq{SM} can be parameterized by the detuning $\delta$ (recall that the total energy $k_1+k_2 =k'_1+k'_2 \equiv K = 2 \Delta +2 \delta$ is conserved) and the two relative coordinates $d = y_1 - y_2$ and $d'=z_1 - z_2$.

First, we separate elastic and inelastic contributions to \eq{SM}.
Picking the unity in the square brackets in \eq{SM}, we see that the two integrations disentangle, and we recognize the elastic contribution $S_M^{(1)} (y_1 , z_1) S_{M}^{(1)} (y_2 , z_2) + S_M^{(1)} (y_1 , z_2) S_{M}^{(1)} (y_2 , z_1)$ to the two-photon scattering matrix \eq{S2_dec}. The remaining term in the square brackets in \eq{SM} generates the inelastic contribution
\beq
i \mathcal{T}_M^{(2)} (y_1 , y_2 , z_1 , z_2) 
&=& - 2 i \kappa^3 \Theta (z_1 >z_2 >y_2 > y_1) \nonumber\\
&\times& \sum_{a,b}  \frac{C_a C_b}{\Delta_a - \Delta_b + i \kappa }  e^{i (\Delta_a -i \kappa/2) (y_1 - z_1) +i (\Delta_b - i \kappa/2) (y_2 - z_2)} \nonumber \\
& & + (y_1 \leftrightarrow y_2) \cdot (z_1 \leftrightarrow z_2),\label{TM3}
\eeq
where
\beq
C_{a}=\prod_{a'=1; a' \neq
a}^M \frac{\Delta_{a}-\Delta_{a'}-i\kappa}{\Delta_{a}-\Delta_{a'}}.
\label{Ca}
\eeq

The elastic  (reducible) part can be cast to
\beq
S_{M; red}^{(2)} (d, d') &=& \frac{1}{2\pi} \int d k \left[ e^{i k (d-d')  } + e^{- i k (d+d')} \right] \nonumber\\
&\times&\left(\frac{\delta +k -i \kappa/2}{\delta +k + i \kappa/2} \right)^M  \left(\frac{\delta -k -i \kappa/2}{\delta -k + i \kappa/2} \right)^M  \nonumber \\
&= & \frac{1}{2\pi} \int d k \left[ e^{i k (|d|- |d' |)  } + e^{- i k (|d| +|d' |)} \right] \nonumber\\
&\times&\left(\frac{\delta +k -i \kappa/2}{\delta +k + i \kappa/2} \right)^M  \left(\frac{\delta -k -i \kappa/2}{\delta -k + i \kappa/2} \right)^M .
\eeq
Evaluating this integral in terms of the $M$th order residua, we obtain the contributions \eq{Sdelta} and \eq{Selast}.

The inelastic (irreducible) part \eq{TM3} admits the integral representation
 \beq
i \mathcal{T}_M^{(2)} (y_1 , y_2 , z_1 , z_2) &=&
- 2  \kappa^3  \Theta (z_1 > z_2 > y_2 > y_1) \nonumber \\
& \times & \int_0^{\infty} d \tau e^{- \kappa \tau} \left( \sum_{a} C_a e^{i (\Delta_a - i\kappa/2) (\tau + y_1 -z_1)} \right) \nonumber\\
&\times&\left(\sum_{b} C_b e^{i (\Delta_b - i\kappa/2) (-\tau +y_2 -z_2)} \right) \nonumber \\
&+& (y_1 \leftrightarrow y_2) \cdot (z_1 \leftrightarrow z_2),
\label{TM3s}
\eeq
which helps to decouple the sums over indices $a$ and $b$. Assuming in the following $z_1 > z_2 > y_2 > y_1$ and employing the identities [cf. Eq.~(\ref{ftss})]
\beq
& &\lim_{\Delta_a \to \Delta} \Theta (x) \sum_a C_a e^{-i (\Delta_a - i \kappa/2) x }\nonumber\\ 
&=& - \frac{\Theta (x)}{2 \pi} \int d s e^{- i s x} \sum_{m=1}^M C_m^M \frac{(- i \kappa)^m}{(s- \Delta + i \kappa/2)^m} , \\
& &\lim_{\Delta_a \to \Delta} \Theta (-x) \sum_a C_a e^{-i (\Delta_a + i \kappa/2) x }  \nonumber\\
&=&  - \frac{\Theta (-x)}{2 \pi} \int d s e^{- i s x} \sum_{m=1}^M C_m^M \frac{( i \kappa)^m}{(s- \Delta - i \kappa/2)^m} ,
\eeq
we transform \eq{TM3s} into
\beq
\!\!\!\!\!\!\!\!\!\!\!\!\!i \mathcal{T}_M^{(2)} (y_1 , y_2 , z_1 , z_2) &=&- 2  \kappa  \int \frac{d s }{2 \pi}\int \frac{d r }{2 \pi} \int_0^{z_1 -y_1-0^+}\!\!\!\!\!\!\!\! d \tau e^{- i (r -s - i\kappa) \tau-is (z_1 - y_1 )-i r (z_2 - y_2)} \nonumber \\
&  \times & \sum_{m_1 =1}^M \sum_{m_2 =1}^M C_{m_1}^M C_{m_2}^M\frac{ (- i \kappa)^{m_1}}{(s - \Delta + i \kappa/2)^{m_1}}   \frac{ (- i \kappa)^{m_2}}{(r - \Delta + i \kappa/2)^{m_2}}
 \nonumber \\
&+& 2  \kappa  \int \frac{d s }{2 \pi}\int \frac{d r }{2 \pi} \int_{z_1 - y_1 +0^+}^{\infty}\!\!\!\!\!\!\!\!\!\!d \tau e^{-i(s+r - i 0^+ ) \tau- i (s +i \kappa) (y_1 -z_1)- i r (z_2 - y_2 )}\nonumber \\
 &\times  &\sum_{m_1 =1}^M \sum_{m_2 =1}^M C_{m_1}^M C_{m_2}^M \frac{(i \kappa)^{m_1}}{(s + \Delta + i \kappa/2 )^{m_1}}    \frac{ (- i \kappa)^{m_2}}{(r - \Delta + i \kappa/2)^{m_2}} .\nonumber\\
\eeq
Integrating over $\tau$ we obtain
\beq
i \mathcal{T}_M^{(2)} (y_1 , y_2 , z_1 , z_2)
&=&  2  i \kappa \int \frac{d s }{2 \pi}\int \frac{d r }{2 \pi}  \frac{e^{- i s (z_1 - y_1 )} e^{- i r (z_2 - y_2)} }{r -s - i \kappa } \nonumber \\
&\times& \sum_{m_1 =1}^M \sum_{m_2 =1}^M C_{m_1}^M C_{m_2}^M\frac{ (- i \kappa)^{m_1}}{(s - \Delta + i \kappa/2)^{m_1}}   \frac{(- i \kappa)^{m_2}}{(r - \Delta + i \kappa/2)^{m_2}} \nonumber \\
&-& 2  i \kappa \int \frac{d s }{2 \pi}\int \frac{d r }{2 \pi} \frac{e^{ -\kappa (z_1 -y_1)}  e^{- i r (z_1 - y_1 + z_2 - y_2 )}}{s+r - i 0^+} \\
&\times&\sum_{m_1 =1}^M \sum_{m_2 =1}^M C_{m_1}^M C_{m_2}^M \frac{(i \kappa)^{m_1}}{(s + \Delta + i \kappa/2 )^{m_1}}    \frac{ (- i \kappa)^{m_2}}{(r - \Delta + i \kappa/2)^{m_2}} .\nonumber
\label{TMinteger}
\eeq
The second term exactly compensates the contribution from the pole $s =r - i \kappa$ in the first term. So we can rewrite \eq{TMinteger} as 
\beq
i \mathcal{T}_M^{(2)} (y_1 , y_2 , z_1 , z_2)
&=&  - 2  i\kappa \int_{\gamma'_1} \frac{d s }{2 \pi}\int_{\gamma'_2} \frac{d r }{2 \pi}  \frac{e^{- i s (z_1 - y_1 )} e^{- i r (z_2 - y_2)}}{s -r + i \kappa }  \\
 &\times&\sum_{m_1 =1}^M \sum_{m_2 =1}^M C_{m_1}^M C_{m_2}^M\frac{ (- i \kappa)^{m_1}}{(s - \Delta + i \kappa/2)^{m_1}}   \frac{(- i \kappa)^{m_2}}{(r - \Delta + i \kappa/2)^{m_2}},\nonumber
\label{tm2mod}
\eeq
where the contours of integration are deformed to the small circles $\gamma'_1$ and $\gamma'_2$ embracing clockwise the poles $s=\Delta- i \kappa/2$ and $r= \Delta - i \kappa/2$, respectively.

Transforming \eq{tm2mod} plus its complements $(y_1 \leftrightarrow y_2) \cdot (z_1 \leftrightarrow z_2)$ to the mixed representation of the total energy $K=2 \Delta +2 \delta$ and the relative coordinates $d,d'$, we obtain
\beq
i \mathcal{T}_M^{(2)} (d,d')
&=&   2  \kappa \int_{\gamma'_1} \frac{d s }{2 \pi}\int_{\gamma'_2} \frac{d r }{2 \pi} \frac{ e^{ i (\Delta + \delta -s)  (|d|+|d'|)} }{(s -r + i \kappa) (2 \Delta +2 \delta -  s -  r )  }  \nonumber \\
 &\times&\sum_{m_1 =1}^M \sum_{m_2 =1}^M C_{m_1}^M C_{m_2}^M\frac{ (- i \kappa)^{m_1}}{(s - \Delta + i \kappa/2)^{m_1}}   \frac{(- i \kappa)^{m_2}}{(r - \Delta + i \kappa/2)^{m_2}}
\label{tm2mod2} \\
&=&  2   \kappa  \int_{\gamma''_1} \frac{d s }{2 \pi}\int_{\gamma''_2} \frac{d r }{2 \pi} \frac{e^{- i (s -\delta - i \kappa/2) D}}{(s -r + i \kappa) (2 \delta -s-r +i \kappa) } \nonumber \\
&\times&\sum_{m_1 =1}^M \sum_{m_2 =1}^M C_{m_1}^M C_{m_2}^M\frac{ (- i \kappa)^{m_1}}{s^{m_1}}   \frac{(- i \kappa)^{m_2}}{r^{m_2}} ,
\label{tm2mod3}
\eeq
where $D=|d| +|d'|$, and the contours $\gamma''_1$ and $\gamma''_2$ embraces the origin clockwise.
Noting that
\beq
& &\sum_{m_1 =1}^M \sum_{m_2 =1}^M C_{m_1}^M C_{m_2}^M\frac{ (- i \kappa)^{m_1}}{s^{m_1}}   \frac{(- i \kappa)^{m_2}}{r^{m_2}} \nonumber\\
&=& \left[ \frac{(s- i \kappa)^M}{s^M}-1\right]\left[ \frac{(r- i \kappa)^M}{r^M}-1\right], 
\eeq
we  cast \eq{tm2mod3} to
\beq
& &i \mathcal{T}_M^{(2)} (d,d')\nonumber\\
&=& - \frac{2  \kappa e^{(i \delta  - \kappa/2) D}}{[(M-1)!]^2} \frac{\partial^{2 M-2}}{\partial s^{M-1} \partial r^{M-1}} \left[ \frac{e^{- i s D} (s- i \kappa)^M (r- i \kappa)^M}{(s -r + i \kappa) (2 \delta -s-r +i \kappa)} \right]_{s=r=0} .
\label{tpp}
\eeq

Representing
\beq
 \frac{(r -i \kappa )^M}{s-r + i \kappa } &=& - (r-i \kappa)^{M-1} \frac{1}{1- \frac{s}{r -i \kappa}} =  - \sum_{l=0}^{\infty} s^l (r-i \kappa)^{M-1-l} \\
&=& - \sum_{l=0}^{M-1} s^l (r-i \kappa)^{M-1-l} - s^M \sum_{l=0}^{\infty} s^l (r-i \kappa)^{-1-l} \nonumber \\
&=& -r^{M-1} + p_{M-2} (r) + \frac{s^M}{s-r+ i \kappa},
\eeq
where $p_{M-2} (r)$ is a polynomial of the degree $M-2$, we establish the identity
\beq
\frac{\partial^{M-1}}{\partial r^{M-1}} \left[ \frac{(r -i \kappa )^M}{s-r + i \kappa } \right]_{r=0} = (M-1)! \left[ \frac{s^M}{(s+ i \kappa)^M} -1 \right],
\eeq
and hence
\beq
& &\frac{\partial^{M-1}}{\partial r^{M-1}} \left[ \frac{(r -i \kappa )^M}{(s-r + i \kappa) (2 \delta -s-r + i \kappa)} \right]_{r=0} \nonumber\\
&=& \frac{(M-1)!}{2 (s - \delta)} \left[ \frac{(s - 2 \delta)^M}{(s - 2 \delta - i \kappa)^M} - \frac{s^M}{(s + i \kappa)^M}\right].
\eeq
With its help we further transform \eq{tpp} and obtain the contribution \eq{Sinelast}.

At $\delta=0$ Eq.~\eq{Sinelast} amounts to
\beq
i \mathcal{T}^{(2)}_M (d,d') &=& - \frac{  \kappa e^{  - \kappa D/2}}{(M-1)!} \frac{\partial^{M-1}}{\partial s^{M-1}} \left\{ e^{-i s D} s^{M-1}  \left[ 1 - \frac{(s-i \kappa )^M}{(s + i \kappa)^M}\right] \right\}_{s=0} \nonumber \\
&=& - 2  \kappa e^{  - \kappa D/2} \frac{1 - (-1)^M}{2},
\eeq
which explicitly proves the parity effect \eq{oddS},\eq{evenS}.

\section*{Acknowledgements} 
We would like to thank D. Baeriswyl, D. Fioretto, and V. Yudson for useful discussions. M.R. and V.G. are supported by MaNEP and
Swiss NSF. M. P. is supported by DFG-FG 723. V.G. thanks KITP for hospitality.


\begin{thebibliography}{99}
\bibitem{mask}
It should be remembered that $|\phi_{1, \mathrm{out}}(x - t)|^2 \approx |\phi_{1,\mathrm{scatt}}(x-t) |^2$ only for $|x-t| \gg \sigma$, since in the region $|x-t|\ll \sigma$ the scattered part is masked by the non-scattered part $\phi_{1, \mathrm{in}}$.
\end{thebibliography}
\end{document}